\documentclass[aps,prb,twocolumn,showpacs]{revtex4}

\usepackage{amsmath}
\usepackage{amssymb}
\usepackage{graphicx}
\usepackage{dcolumn}
\usepackage{bm}

\begin{document}

\title{Giant optical anisotropy in cylindrical 
self-assembled InAs/GaAs quantum rings}

\author{Weiwei Zhang, Zhiqiang Su, Ming Gong, Chuan-Feng Li, Guang-Can Guo,
Lixin He \footnote{corresponding author,
Email address: helx@ustc.edu.cn} 
}
\affiliation{Key Laboratory of Quantum Information,
University of Science and Technology
of China, CAS, Hefei, 230026, People's Republic of China}
\date{\today}

\begin{abstract}

Using a single-particle atomistic pseudopotential method followed by
a many-particle configuration interaction method,
we investigate the geometry, electronic structure and optical transitions of
a self-assembled InAs/GaAs 
quantum ring (QR), changing its shape continously from a lens-shaped 
quantum dot (QD)
to a nearly one dimensional ring. 
%
We find that the biaxial strain in the ring is strongly asymmetric in
the plane perpendicular to the QR growth direction, leading to giant 
optical anisotropy. 

\end{abstract}

\pacs{73.21.La, 73.22.-f, 71.35.-y}





\maketitle

Recently, a novel nano-structure, quantum ring (QR),
has been fabricated via the
self-assemble techniques,\cite{garcia97} in various semiconductor
systems, such as In(Ga)As/GaAs,
\cite{garcia97, huang06,huang07, granados05, gong05} InAs/InP, \cite{raz03, sormunen05}
InP/GaInP, \cite{wipakorn07} and Si/Ge\cite{cui03} etc., with
controlled sizes and shapes. 
Like the self-assembled quantum dots, the QRs have
discrete energy levels due to the 3D confinement
effects.
However, a QR differs from a QD because
of its non-simply connected topology,
and therefore offers a unique opportunity to study the physical effects
in addition to the confinement effects, such as the Aharonov-Bohm
effect,\cite{aharonov59,keyser03} and quantized magnetic
susceptibility,\cite{chakarborty94} etc.

The electronic structure of self-assembled QRs have been explored
via the electron charging experiment \cite{lorke00}
and the photoluminescence (PL) spectra of charged excitons.
\cite{warburton00,haft02}
On the other hand, most of the theoretical studies on the QRs are
still at the continuum theory level, such as the effective mass
approximations
(EMA),\cite{cheung88,chakarborty94,li00,li02,song01,llorens01} ${\bf
k\cdot p}$ method,\cite{planelles02} 
and the local spin density
approximation,\cite{climente06} etc., assuming 1-dimensional,
\cite{cheung88} 2-dimensional,
\cite{chakarborty94,song01,llorens01} and 3-dimensional model
confinement potentials. \cite{li00,planelles02,filikhin06}
These studies provide valuable qualitative
knowledge about the single-particle electronic
structures, \cite{planelles02,llorens01, chakraborty94,filikhin06} 
as well as the many body effects
\cite{climente06,pederiva02,song01} of the QRs.
However, it has been shown that
an atomistic theory is necessary to capture
the subtle features, such as energy level
splittings,\cite{williamson00} shell filling \cite{he05d}
and exciton fine structures \cite{bester03} etc.
at single dot level.
The atomistic effects are expected to be even
more important for the QRs
because the QRs have much larger surface area to volume ratio
than the QDs of similar sizes.
%

\begin{figure}
\begin{center}
\includegraphics[width=2.0in]{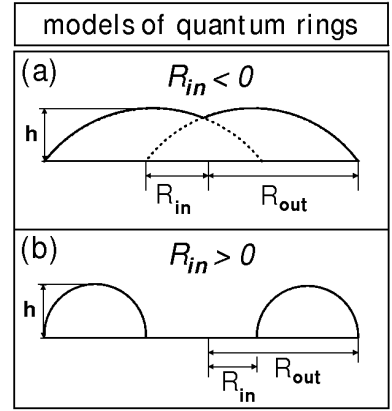}
\end{center}
\caption{Cross sections of the QRs for (a) $R_{in} <$ 0
and (b) $R_{in} >$ 0, where $R_{in}$ and $R_{out}$
are the inner radius and  outer radius of the QRs,
respectively. The QR heights $h$ are fixed to be 2.5 nm.
}
\label{fig:ringprofile}
\end{figure}

In this paper, we investigate the electronic structures and optical
transitions of
realistic self-assembled InAs/GaAs QRs via
an atomistic pseudopotential method.\cite{wang99b}
We change the geometries of QRs from a lens-shaped QD to a narrow ring, by
continously increasing the inner radius of the ring.
We then study the strain profiles,
the single-particle energy levels,
as well as the optical transitions of the QRs 
with respect to the inner radius.
The QRs have more complicate strain profiles
than the QDs, due to their complicate
topology. We show that the biaxial strain of QRs is strongly asymmetric in the
plane perpendicular to the QR growth direction, 
leading to single-particle energy level
crossing and giant optical anisotropy (even in cylindrical QRs).

\begin{figure}
\begin{center}
\includegraphics[width=3.0in]{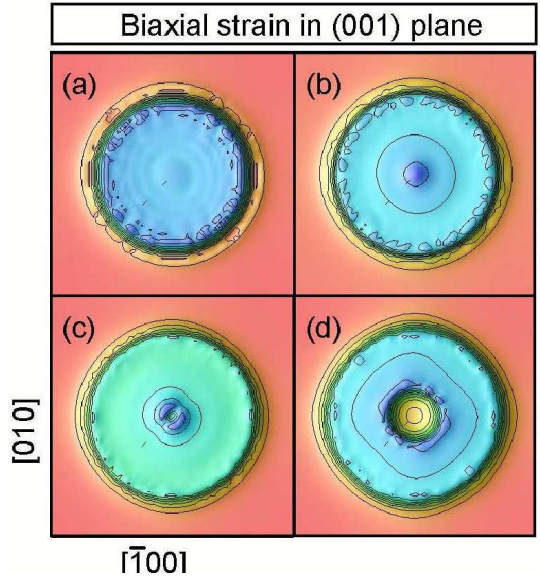}
\end{center}
\caption{(Color online) The biaxial strain in the (001) plane for
(a) $R_{in}=-R_{out}$ (lens-shaped QD);
(b) $R_{in}$=-3 nm;
(c) $R_{in}$=0 nm;
and (d) $R_{in}$=3 nm.
}
\label{fig:biaxial}
\end{figure}


Figure \ref{fig:ringprofile} depicts
the cross sections of the QRs with their structural
parameters, embedded in a
60$\times$60$\times$60 GaAs matrix. The QRs are assumed
growing along
the [001] direction, on the top of a 1-monolayer wetting layer.
The outer radius $R_{out}$ of the QR is measured from the center of
the base to the outside edge of the ring, whereas the inner radius
$R_{in}$ is defined to be the distance from center of the base to
the inner circle. At $R_{in}=-R_{out}$, the QR is a lens-shaped
quantum dot.  \cite{ring}
Therefore, by increasing the 
inner radius, we continously change the QR
from a lens-shaped
quantum dot to a one-dimensional quantum wire 
as $R_{in} \rightarrow R_{out}$.
We fix the height of the quantum ring $h$= 2.5 nm, 
outer radius $R_{out}$=12.5 nm. 
and vary the inner radius $R_{in}$ from -12.5 nm (lens-shaped dot) to 6 nm.
Alloy QRs and rings of larger outer radii give very similar results.

The single-particle energy levels
and wavefunctions of the rings are obtained 
by solving the Schr\"{o}dinger equations,
\begin{equation}
\left[ -{1 \over 2} \nabla^2
 + V_{\rm ps}({\bf r}) \right] \psi_i({\bf r})
=\epsilon_i \;\psi_i({\bf r}) \; ,
\label{eq:schrodinger}
\end{equation}
where the total electron-ion potential $ V_{\rm ps}({\bf r})$
is a superposition of
local, screened atomic pseudopotentials $v_{\alpha}({\bf r})$,
\cite{williamson00}
i.e.
$V_{\rm ps}({\bf r}) =\sum_{n,\alpha}
v_{\alpha}({\bf r} - {\bf R}_{n,\alpha})$.
The atom positions $\{ {\bf R}_{n,\alpha} \}$
are obtained by
minimizing the strain energies using the valence force
field (VFF) method. \cite{keating66,martins84}
Equation~(\ref{eq:schrodinger}) is solved using the ``Linear
Combination of Bloch Bands'' (LCBB) method. \cite{wang99b} 
The exciton energies and optical transitions
are calculated via a 
configuration interaction (CI) method, \cite{franceschetti99}
in which the exciton wavefunctions are expanded as the linear
combination of Slater determinants constructed from the single-particle 
electron and hole wavefunctions.

{\it Strain profiles}:  
Figure \ref{fig:biaxial} (a-d) depict the biaxial strain,
\begin{equation}
{\rm B=}\sqrt{(\epsilon_{xx}-\epsilon_{yy})^2+
(\epsilon_{zz}-\epsilon_{xx})^2+(\epsilon_{yy}-\epsilon_{zz})^2}\, ,
\end{equation}
of the QR, for $R_{in}$=-12.5 (lens-shaped dot), -3, 0 and 3 nm,
respectively.
For the lens-shaped dot,
the biaxial strain is almost isotropic in the (001) plane.
However, with the increasing of the inner radius $R_{in}$,
the biaxial strain becomes asymmetric
in the (001) plane:
the biaxial strain along the [110] direction
becomes larger than that along the
[1$\bar{1}$0] direction. As the inner radius
increases further, the difference of biaxial strain 
between the two directions becomes significant.
One clearly see
the biaxial strain peaks around the inner edge of the ring
along the [110] direction
at $R_{in}$=0 nm and $R_{in}$=3 nm
in Fig. \ref{fig:biaxial}(c),(d).
Since the hole confinement potential strongly depend on the biaxial strain, 
as a consequence, the strain-modified potentials for holes (not shown) are
also asymmetric in the (001) plane.

\begin{figure}
\includegraphics[width=3.8in,angle=-90]{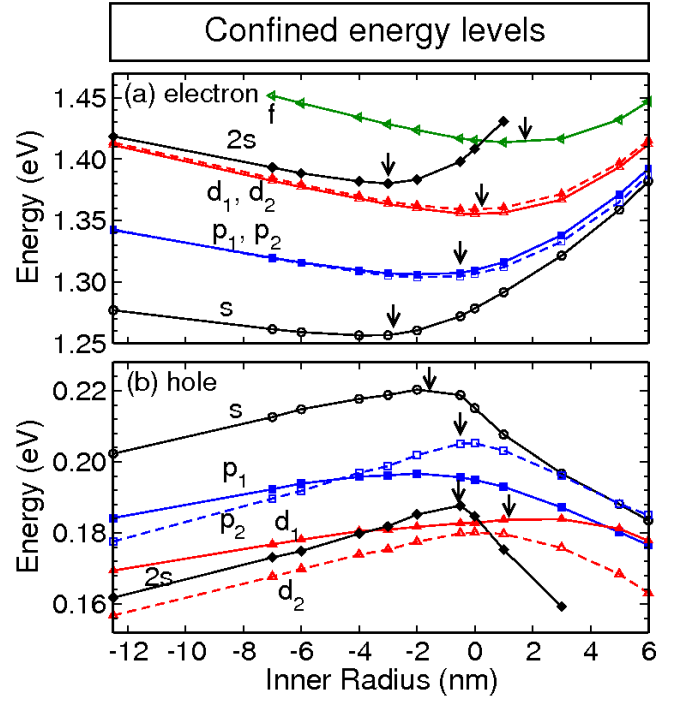}
\caption{(Color online) The single-particle energy levels of
confined (a) electron states and (b) hole state.
The reference energy is choose to be the valance band edge of GaAs. }
\label{fig:ringe}
\end{figure}

\begin{figure}
\begin{center}
\includegraphics[width=3.5in]{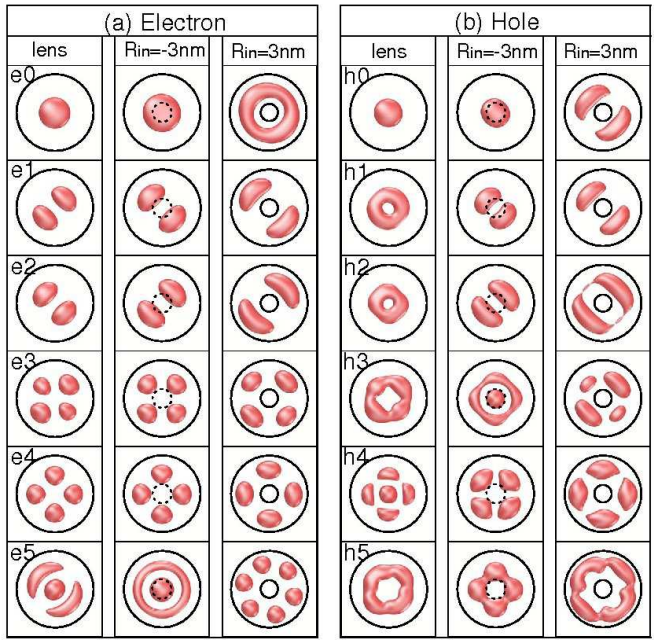}
\end{center}
\caption{(Color online) Top view
of the squared wavefunctions of the confined (a) electron and (b) hole states
in the QRs. We show the wavefunctions at $R_{in}$= -12.5, -3 , 3 nm, respectively.
The iso-surface is chosen to enclose 50\%
of the density of the state.
The crystallographic directions are same as in Fig.\ref{fig:biaxial}.
}
\label{fig:wavefunction}
\end{figure}

{\it Single-particle energy levels and wavefunctions}:
The electron and hole single-particle energy levels are shown in
Fig. \ref{fig:ringe} (a),(b) respectively,
as functions of $R_{in}$. The orbital labels $s$, $p$, etc.
are given by continuously monitoring the changing
characters of the wavefunctions (see Fig. \ref{fig:wavefunction}).
We denote the $p$, $d$ orbitals with peaks along the
[1$\bar{1}$0] direction $p_1$, $d_1$ and 
those of peaks along the
[110] direction $p_2$, $d_2$. 
For lens-shaped dot, the $s$-$p$ and $p$-$d$ energy spacings are nearly
equal.
For electrons, the two $p$ orbitals are nearly degenerate, as well as the two
$d$ levels. However, for holes, the $p$, $d$ levels show quite large 
(8 - 10 meV) energy splitting.~\cite{williamson00,he06a} 
The 2$s$ state is close in energy to the $d$ states, suggesting that the
confinement potential is close to a parabolic potential. \cite{footnote1}
As we increase $R_{in}$, we found that all the
confined electron (hole) levels decrease (increase) first and then increase
(decrease).
We attribute the decreasing (increasing) of the energy levels at small $R_{in}$
to the strain effects after a close examination of the
strained confinement potentials.
The energy level turning points are
indicated by the small arrows in Fig.\ref{fig:ringe}.
The turning points $R_{in}$ of the $s$ and 2$s$ levels,
are about -2 to -3 nm, whereas
the turning points of $p$ states are at larger inner radius
$R_{in}$=-1 nm. The turning points of the $d$ states are at
about $R_{in}$=1 nm for both electrons and holes, larger than those of $s$ and
$p$ orbitals.
After the turning point, the confined energy levels increase rapidly with
the increasing of $R_{in}$, especially for the $s$ and $2s$ states. The 2$s$
states become unconfined at $R_{in} \sim $ 2 nm.

The trend of the electron energy
levels as functions of the $R_{in}$ of the QRs can be understood by
examining the wavefunctions of each level, shown in
Fig. \ref{fig:wavefunction}(a, b) for the
six lowest confined electron and hole states.
For lens-shaped QDs, the electron wavefunctions are the
$s$ ($e_0$), $p$ ($e_1$, $e_2$), $d$ ($e_3$, $e_4$)
and 2$s$ ($e_5$) orbitals respectively,
in the order of increasing energy, whereas for holes, the 2$s$ orbital is
between the two $d$ levels.
The character that distinguishes 
the 2$s$ orbital from the $d$ orbitals
is that the 2$s$ orbital has a maximum at the center of the dot, 
whereas the $d$ orbitals have nodes at the dot center.
The electron 2$s$ orbital becomes unconfined after $R_{in} >$0 nm, 
and $e_5$ is actually a $f$ state at $R_{in}=$3 nm.
The $s$, 2$s$ states, having maximum density at the dot center,
are most sensitive to $R_{in}$, 
as the states feel strong confinement potential from the
inner circle of the ring with increasing of $R_{in}$, 
leading to increasing of the level energy much 
faster than other states. Their wavefunctions 
also change dramatically from disk-like states at
$R_{in} <$0 nm (i.e., has the maximum at center of the ring), to
ring-like (i.e., hollow at the center of the ring) at
$R_{in} >$0 nm.
In contrast, the $p$ and $d$ orbitals have nodes at the ring center,
and do not feel the confinement potential until at much
larger $R_{in}$.
As a consequence, the electronic structures of the QRs
deviate significantly from those of the QDs:
(i) The $s$-$p$ energy level spacing is much smaller than the $p$-$d$ energy
level spacing;
(ii) The 2$s$ level is no longer (nearly) degenerate with the two $d$ levels.

We observe several energy level crossings with respect to $R_{in}$ in
Fig. \ref{fig:ringe}, including
the level crossing between electron 2$s$ state and $f$ state at $R_{in}$=0 nm, 
the hole 2$s$ state with two $d$ states at $R_{in}$= -3 and 0 nm, as well as
the two electron (hole) $p$ orbital at $R_{in}$= -3 (-4) nm.
The level crossing of different angular momentum, (e.g. the 2$s$ and $f$
states) is due to the confinement effect, as discussed above,
whereas the level crossing between the two hole $p_1$ and $p_2$ states
is due to the biaxial strain effects.

\begin{figure}
\begin{center}
\includegraphics[width=3.8in, angle=-90]{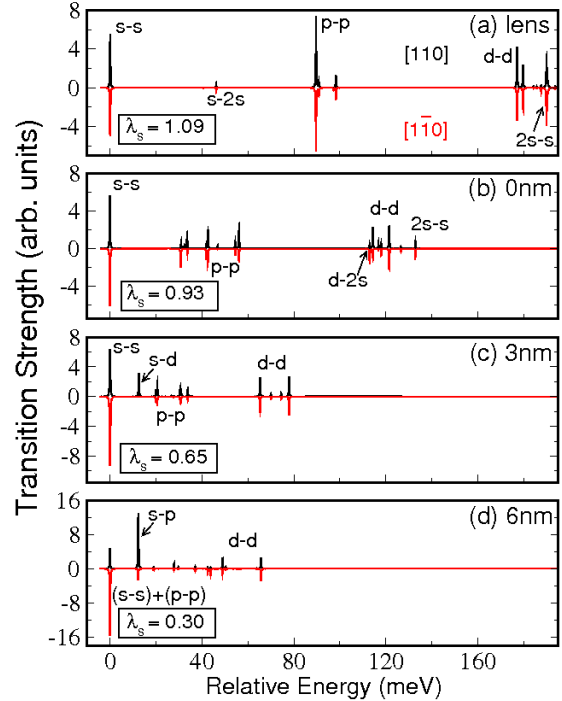}
\end{center}
\caption{(Color online) The single exciton absorption spectrum for QRs
with different inner radius. 
The primary exciton energy at each panel is shifted to zero.} 
\label{fig:spectrum}
\end{figure}

{\it Optical transitions}:
We calculate the optical transitions spectrum in QRs, and the results are
shown in Fig. \ref{fig:spectrum} (a-d) 
for $R_{in}$= -12.5, 0, 3, 6 nm, 
respectively.
We show the polarized transition intensities in both [110] 
and [1$\bar{1}$0] directions, as the transitions are 
almost linearly polarized 
in these two directions. 
The primary exciton energy is shifted to zero for each $R_{in}$ for clarity, 
and the transition peaks are marked by their leading 
transition characters. 
For example, the primary exciton transition is a $s$ (electron)
to $s$ (hole) transition. We show also the $p$-$p$ and $d$-$d$
transitions. 
As we see, the energy differences
between the $s$-$s$, $p$-$p$ and $d$-$d$ transitions 
decrease significantly with 
the increasing of inner radius, reflecting the
change of single particle level spacings. 
The transition intensities also change dramatically with respect to
$R_{in}$. For example, the $p$-$p$ and $d$-$d$ transitions are much weaker in
the QRs than in the dots. 
At $R_{in}$= 6 nm, a significant $s$ to $p$ transition appears, because 
the envelope wavefunctions of electron $s$ state and hole $p$ states 
are no longer (nearly) orthogonal at this $R_{in}$.

Interestingly, even though the total transition intensity of primitive exciton
does not change much
with respect to $R_{in}$, the transition intensity polarized along the
[1$\bar{1}$0]  becomes much stronger than that of the 
[110] direction. 
We calculate the optical polarization anisotropy $\lambda$, 
defined as the ratio of the transition intensities along
the [110] and $[1\bar{1}$0] direction, i.e., \cite{williamson00}
\begin{equation}
\lambda ={I_{[110]} \over I_{[1\bar{1}0]}} \, ,
\label{eq:lambda}
\end{equation}
for the $s$-$s$ transitions.
For lens-shaped QDs [Fig. \ref{fig:spectrum} (a)],
$I_{[1\bar{1}0]}$ is slightly smaller than
$I_{[110]}$ ($\lambda_s$=1.09), 
agree with previous calculations.\cite{williamson00} 
However, with increasing of the inner radius,
$\lambda_s$ decrease dramatically. At $R_{in}$= 6 nm, 
$\lambda_s$ is only about 0.30, meaning the transition in the [1$\bar{1}$0]
direction is about 3 times stronger than in the [110] direction. 
Since the QRs studied here are cylindrical, the
giant optical anisotropy is because of the ``atomic symmetry
factor'' \cite{williamson00} due to the 
asymmetric biaxial strain.

To conclude, we have investigated via a single-particle atomistic
pseudopotential and
a many-particle CI methods, the electronic structures and optical transitions
of self-assembled InAs/GaAs QRs. We find that even in cylindrical 
InAs/GaAs quantum rings, the biaxial strain is strongly
asymmetric in the (001) plane, where
the biaxial strain along the [110] direction is much larger than that
along the [1$\bar{1}$0] direction. The asymmetric strain induces
single-particle energy level crossing,
and lead to giant optical anisotropy. 
The optical anisotropy can be examined 
in future experiments and should be taken account of in designing
QR devices.

L.H. acknowledges the support from the Chinese National
Fundamental Research Program 2006CB921900, the Innovation
funds and ``Hundreds of Talents'' program from Chinese Academy of
Sciences, and National Natural Science Foundation of China (Grant
No. 10674124).



\begin{thebibliography}{35}
\bibitem[{rin()}]{ring}
\bibinfo{note}{Rigiously speaking, when $R_{in}<0$, the QR is not a {\it real}
  ring. However, to simplify the discussion, we still call it a ring in this
  paper.}
\bibitem[{foo()}]{footnote1}
\bibinfo{note}{For a cylindrical QDs, the degeneracy of energy levels is at
  most two. However, for a dot with parabolic confinement, the 2$s$ level also
  {\it accidentally} degenerate with the $d$ levels.}
\end{thebibliography}
\end{document}